# Scaling behaviour of the helical and skyrmion phases of $Cu_2OSeO_3$ determined by single crystal small angle neutron scattering

J. Sauceda Flores,[1] R. Rov,[2] J. O'Brien,[1] S. Yick,[1,2,3,4] Md. F. Pervez,[1] M. Spasovski,[2] J. Vella,[2] N. Booth,[4] E. P. Gilbert,[4] Oleg A. Tretiakov,[1] T. Söhnel[2,3,†] and C. Ulrich[1,*]

[1] *School of Physics, The University of New South Wales, NSW 2052, Australia*
[2] *School of Chemical Sciences, University of Auckland, Auckland 1142, New Zealand*
[3] *MacDiarmid Inst. Adv. Mat. & Nanotechnology, Wellington 6140, New Zealand*
[4] *Australian Centre for Neutron Scattering, Australian Nuclear Science and Technology Organisation, Lucas Heights NSW 2234, Australia*

(Dated: 4. March 2023)

**ABSTRACT**

Skyrmions are topologically protected quantum objects at the nanometre scale. They form perpendicular to an applied magnetic field at a certain temperature and arrange themselves in a typically hexagonal lattice. Using small angle neutron scattering we have determined the magnetic field versus temperature phase diagrams of the stability range of the different magnetic phases, the helical as well as the skyrmion phases in the multiferroic skyrmion material $Cu_2OSeO_3$. Therefore, a single crystal was mounted in different crystal orientations, i.e. unoriented and with the crystallographic ⟨1 1 0⟩ and ⟨1 0 0⟩ axis aligned parallel to the magnetic field. Furthermore, different cooling procedures were tested, cooling from the paramagnetic phase at zero magnetic field and field cooling through the skyrmion phase where metastable skyrmions are nucleated. From this, not only were the stability ranges of both the helical and skyrmion phases in this multiferroic skyrmion material determined, but the length of the spin helix and the skyrmion distances at different conditions was also determined through detailed analysis of the positions of the observed Bragg peaks. The obtained data provide valuable information about the scaling of the skyrmion distances and therefore their packing density. The knowledge about this tuneability will serve as an important input for the theoretical understanding of the formation of skyrmions. Concerning technological applications, for example as high-density data storage devices in information technology, the temperature and magnetic field dependence of the packing density of skyrmions is of critical importance for their design. Therefore, the obtained knowledge is an essential input for future skyrmionics applications in high-density data storage, in low-energy spintronics or in skyrmionics quantum computation.

# INTRODUCTION

Unconventional topological spin structures such as in chiral spin systems and, in particular, in skyrmion materials offer a plethora of fascinating phenomena for fundamental research and future technological applications [1-4]. A skyrmion is a topologically stable particle-like object comprised of spins forming vortexes at the nanometre scale. These spin rotations are about 10-100 nm in diameter and order in a two-dimensional, typically hexagonal superstructure perpendicular to an applied external magnetic field [5-16]. Their dynamics have links to flux line vortices as in high temperature superconductors. Most skyrmionic systems discovered thus far, such as MnSi are metallic. However, $Cu_2OSeO_3$ (COSO) is a unique case of a multiferroic material where the skyrmion dynamics can be controlled through the application of an external electric field [17-19]. The direct manipulation of the skyrmions through a non-dissipative method offers technological benefits [2,3,19-22] and unique possibilities for testing fundamental theories also related to the Higgs Boson whose theoretical description has similarities to skyrmions.

The insulating skyrmion materials COSO crystallizes in the cubic space group $P2_13$ which consists of groups of edge- and corner-sharing $CuO_5$ trigonal bipyramids (three per group) and square pyramids (one per group), where the trigonal bipyramids are arranged in a diamond type structure [23-25]. The $Cu^{2+}$ ions possess a spin of ½, which are aligned ferromagnetically for the neighbouring square pyramids and antiferromagnetically to the trigonal bipyramids. This leads to a predominantly three up and one down ferrimagnetic structure [26-28]. Due to the complex interplay between the different ferromagnetic and antiferromagnetic interactions together with the non-collinear Dzyaloshinskii-Moriya interaction, the spin structure at the ground state is helical below a critical temperature of about 59 K [29-33]. Upon the application of a weak magnetic field of about 40 mT at base temperature the helical spin structure becomes conical, which can be regarded as a canted spin helix in the direction of the applied magnetic field. As the strength of the magnetic field increases, the spins become field-polarized above about 120 mT. Remarkably, in the temperature range between 55 K and 59 K a skyrmion lattice forms at a finite applied magnetic field of up to 40 mT [7-15]. This indicates that thermal fluctuations play an important role in the formation of this high-temperature skyrmion lattice (HT-SkL). Chacon *et al.* have discovered that a second low-temperature SkL phase (LT-SkL) is formed at temperatures below about 35 K and at higher magnetic fields (about 55 mT to 95 mT) as compared to the HT-SkL phase. This second skyrmion phase is only established when the magnetic field is applied along the cubic ⟨1 0 0⟩ axis.

Upon cooling from the HT-SkL phase, metastable skyrmions can be created which persist far into the conical and helical phase down to lowest temperatures. Their lifetime depends on the final temperature and it is interesting to note that their lifetime ranges from minutes to several hours [34-38]. This can be explained in terms of an energy barrier required for the formation of skyrmions. Once formed in the HT-SkL phase, the topologically stabilized skyrmions remain in a metastable energy minimum upon cooling down to the helical/conical phase. Their decay can be understood as a quantum mechanical tunnelling through this energy barrier. In their small angle neutron scattering experiment, Banneberg *et al.* determined the existence range of metastable skyrmions in the magnetic field - temperature (H-T) phase diagram [15]. Therefore, a COSO single crystal was field cooling at 14 mT, i.e. through the HT-SkL phase. Using this procedure, metastable skyrmions were generated, which were extremely robust and existed down to the lowest temperature.

Concerning technological applications, for example as racetrack memory devices in information technology and for logic gates in spintronics or quantum computation [2,3,19-22], the ability to manipulate skyrmions is of fundamental importance. In general, it has been demonstrated that skyrmions can be created, annihilated, externally detected and moved. Therefore, in device applications, the size and the distance between the skyrmions is an important information since these parameters determine their packing density. Furthermore, the investigation of the skyrmion lattice and helimagnetic structures at equivalent temperatures will provide further insight into the underlying mechanism(s) at play.

In this work we have performed a systematic investigation of the scaling behaviour of the skyrmion distances and the length of the helical propagation vector as a function of magnetic field and temperature throughout the entire H-T phase diagram for different crystal orientations and for different cooling procedures, zero field cooled (ZFC) and field cooled (FC) through the HT-SkL phase. Small Angle Neutron Scattering (SANS) is an ideal technique for the investigation of skyrmion lattices since neutrons are a highly sensitive probe of magnetic behaviour due to their intrinsic magnetic moment and since the distance between the individual skyrmions is between 10-100 nm [5,10-15]. In addition to determining the type of the magnetic structure, i.e. helical or skyrmion lattice, the direction and propagation length of the spin helix or the orientation of the skyrmion lattice and the skyrmion distances can be determined with great accuracy. Our results provide a detailed overview of the scaling behaviour of the skyrmion distances in the H-T phase diagram. The information about this tuneability will provide an important input for the design of future technological devices for example as low-energy consumption, high-density data storage devices. Furthermore, the obtained data will contribute for a further understanding of the creation process and stability of these fascinating topologically protected quantum objects.

**RESULTS**

Neutron diffraction patterns were measured for various crystal orientations, i.e. for the unoriented single crystal and with the ⟨1 0 0⟩ and ⟨1 1 0⟩ crystal axis along the magnetic field, which was parallel to the incoming neutron beam in this experiment. The corresponding neutron Laue diffraction patterns obtained using the instrument JOEY at ANSTO for sample alignment are shown in the Supplementary Information (Figs. 1b-d). Concerning SANS, with measurements being conducted on the instrument QUOKKA at ANSTO, in this geometry the helical spin structure possesses two (or four in case of magnetic twinning) Bragg reflections and the skyrmion phase manifests itself with six Bragg reflections. Note that the conical phase is not visible in this sample geometry [13]. The obtained phase diagram as a function of magnetic field and temperature for the unoriented COSO single crystal single crystal is shown in Figure 1. The data were taken after zero field cooling (ZFC) where the sample was cooled without a magnetic field from 80 K, i.e. from the paramagnetic phase to the lowest possible temperature and the magnetic field was then increased to the desired value. In this way the sample was not cooled through the skyrmion phase and metastable skyrmions were not nucleated. The data were then measured upon heating the sample at the predetermined magnetic field in small temperature steps. The illumination time per diffraction pattern was 1 minute and about 500 diffraction patterns were taken for this phase diagram. Selected raw data are shown in the Supplementary Information (Fig. S2). The intensity phase diagram is shown in Fig. 1.a) and 1.b) where the helical phase appears below $58.4 \pm 0.15$ K at zero magnetic field and the skyrmion phase has an extension from $54.2 \pm 0.15$ K at 30 mT to $58.2 \pm 0.15$ K below 25 mT. Note, the neutron scattering intensity was obtained by integrating over all first order helical or skyrmion Bragg peaks.

To analyse the data, we used a custom numerical routine written in MATLAB which provided the circularly integrated intensity for a selected region of the detector ($6.5 \times 10^{-3}$ Å$^{-1}$ < $|q|$ < $12.5 \times 10^{-3}$ Å$^{-1}$). The Bragg reflections of the helical and skyrmion lattice were manifested as sharp peaks, which were then analysed by fitting Gaussian lineshapes to the experimental data (see Fig. 3). The obtained |q|-values correspond to the length of the helical propagation vector and to the reciprocal vector of the skyrmion lattice. They are plotted in Fig. 1.c) as a function of magnetic field and temperature. For the skyrmion lattice, the largest value of |q| is reached in the centre of the skyrmion phase and decreases when reaching the temperature and magnetic field boundaries of the skyrmion range towards the paramagnetic state on one side or the conical phase on the other side. This corresponds to an extension of the skyrmion distances. Figure 1 d)-f) shows the magnetic field dependence at T = 57 K of the Bragg peak intensity, the absolute value of |q| and the length λ of a spin helix or skyrmion distances in the skyrmion lattice, respectively. At this temperature the skyrmion distance reaches a minimum of 60.9 ± 0.3 nm at 20 mT. In comparison, the length of the spin helix increases at this temperature from 60.1 ± 0.4 nm to 64.7 ± 0.7 nm with increasing magnetic field. This is consistent with our previous Lorentz phase TEM investigation [16]. It is interesting to note that the |q| phase diagram presented in Fig. 1.c) indicates that the largest value of |q| for the helical propagation vector is reached at about 55 K at zero magnetic field and not at the lowest temperature.

In general, the magnitude of the q-vector is a result of the competition between the magnetic exchange interaction J and the Dzyaloshinskii-Morya interaction (DMI) with |q| ~ D/J [4,16,39]. A systematic scaling of the length of the helical propagation vector with magnetic field has been observed before in a thinned COSO single crystal [16] and a temperature dependent increase of the helical propagation vector was determined in cubic $SrFeO_3$ and tetragonal $SrFeO_{2.87}$ single crystals [40], in Fe-films grown on Ir(111) substrate [41], and of the length of the spin cycloid in multiferroic $BiFeO_3$ thin films [42] as well as single crystals [43]. In all these cases the competition between the various magnetic exchange interactions causes a change to the |q|-value, either between long-range and short-range superexchange and double exchange interactions as in the case of $SrFeO_{3-\delta}$, or between collinear magnetic exchange interactions and the DMI.

The magnetic field - temperature (H-T) phase diagram of the ⟨1 0 0⟩ oriented COSO single crystal as measured by SANS is shown in Fig. 2 and selected diffraction patterns are shown in Fig. 3. The data in Fig. 2.a) - c) were measured after ZFC and about 700 diffraction patterns were recorded. The helical phase has a phase transition temperature of 57.1 ± 0.15 K at zero magnetic field and the high-temperature skyrmion phase appears between 51.0 ± 0.15 K and 57.0 ± 0.15 K at 25 mT. A direct comparison of the phase diagrams of the unoriented and ⟨1 0 0⟩ oriented COSO single crystal is shown in the Supplemental Fig. S3. For the ⟨1 0 0⟩ orientated crystal the boundary of the high temperature skyrmion to the paramagnetic phase is lowered by 1.2 K and the low temperature boundary of the HT-SkL phase to the conical phase is reduced by 3.2 K. The HT-SkL phase of the ⟨1 0 0⟩ oriented crystal is expanded as compared to the unoriented COSO single crystal. This is a consequence of the cubic anisotropy term [13], which affects this phase. Finally, it should be noted that the overall intensity of the Bragg reflections is about 10 times stronger for the oriented crystal as compared to the unoriented crystal. Therefore, higher order reflections for both the helical phase and the skyrmion phase are also observed (see Fig. 3). This is most pronounced for the helical phase, but also visible for the HT-SkL phase. Reim *et al.* have attributed the appearance of higher-order reflections to a superlattice which originated from weak modulations of the original magnetic structures [44].

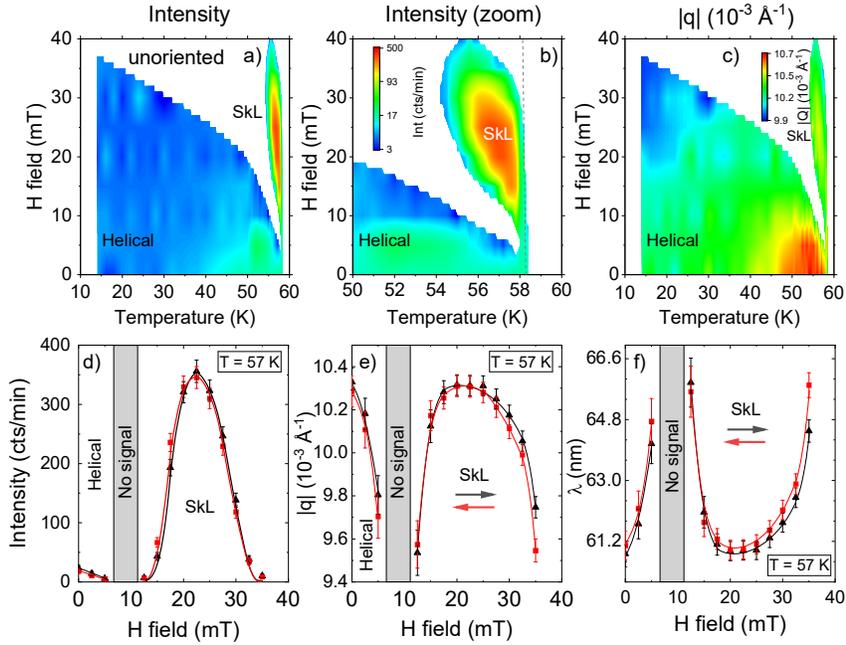

***Fig. 1:*** *a) and b) Changes in the neutron scattering intensity of the unoriented $Cu_2OSeO_3$ single crystal as a function of temperature and magnetic field as obtained by SANS. The helical and the skyrmion phase are clearly visible while the conical phase cannot be seen in this experimental geometry. Note, the intensity is plotted on a logarithmic scale. c) shows the H-T phase diagram of the absolute value of |q|, which corresponds to the magnitude of the helical propagation vector or the reciprocal vector of the skyrmion lattice. The magnetic field dependence at T = 57 K is shown in d) for the integrated Bragg peak intensity, e) the absolute value of |q| and f) the length λ of a spin helix or skyrmion distances in the skyrmion lattice, respectively. The error bars correspond to one standard deviation (s. d.).*

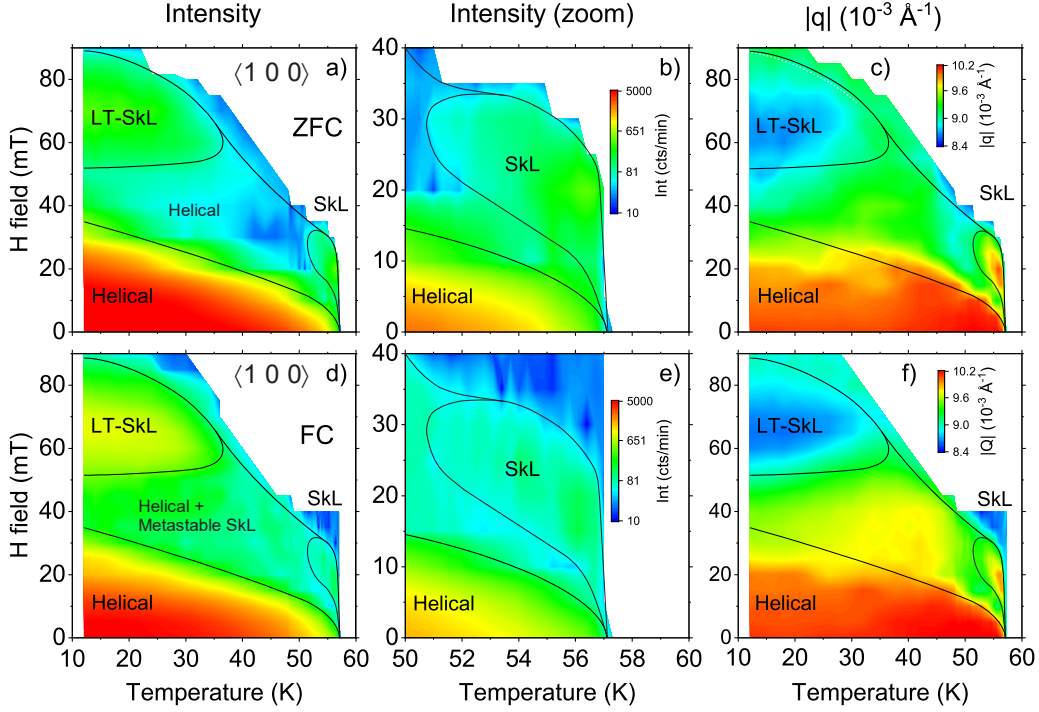

*Fig. 2:* H-T phase diagram of the ⟨1 0 0⟩ oriented COSO single crystal. Panels a)-c) show the data taken after ZFC and d)-f) the data taken after FC at a magnetic field of 20 mT, i.e. cooling through the high-temperature skyrmion phase. The contour maps of the neutron scattering intensities are shown in a) and d) and are plotted on a logarithmic scale. b) and e) show a zoomed in region around the high temperature skyrmion range. In c) and f) the absolute values of the $|q|$-vector of the helical and skyrmion Bragg peaks are plotted. The black solid lines were determined from the changes in the diffraction patterns and serve as guides to the eye.

The most striking difference to the phase diagram of the unoriented single crystal is the appearance of a second skyrmion phase at low-temperature (LT-SkL) and at a magnetic field of about 55 mT to 90 mT. This low-temperature skyrmion phase was first observed by Chacon *et al.* [13] and its appearance can be energetically explained with the cubic anisotropy term. This phase possesses a rather ring-like structure in the diffraction pattern as compared to the 6/12 fold structure of the HT-SkL phase (see Fig. 3.e and f) which is an indication of strong fluctuations in the LT-SkL phase. The helical phase is clearly present below about 30 mT with a strong Bragg peak intensity, however, in contrast to the unoriented sample, weak signatures of the helical phase appear also for higher magnetic fields up to the region of the LT-SkL phase.

The absolute value of |q| of the observed Bragg peaks is presented in Fig. 2.c. For the HT-SkL phase |q| reaches its maximum right in the centre of the HT-SkL phase and its value is slightly larger as compared to the unoriented COSO crystal. In contrast, the |q|-value is about 20 % smaller for the LT-SkL phase and |q| has a flat local minimum in the centre of the LT-SkL phase. For the HT-SkL phase this result is consistent with the data obtained by Chacon *et al.* [13]. It has been discussed whether the two magnetic skyrmion regions of the H-T phase diagram are thermodynamically connected or not since this could point at the nature of the fundamental quantum processes behind their stabilization. The field- and temperature-dependence of the scattering intensity measured upon ZFC (see Fig. 1 and 2) suggest the two different skyrmion phases are separated in the H-T phase diagram by the conical phase. However, features of metastable skyrmions persist between both SkL phases in the SANS data for data taken after FC through the HT-skyrmion phase. Nonetheless, the inverse variations in |q| for the two different SkL regions upon applied magnetic field or temperature suggests that the low-temperature skyrmion state is stabilized in a different way, i.e. an additional term in the spin Hamiltonian, namely the magnetic anisotropy along this preferential direction plays an important role [13,15].

The H-T phase diagram of the ⟨1 0 0⟩ oriented crystal was also measured upon field cooling (FC) at 20 mT. Therefore, the sample was cooled from the paramagnetic phase at 80 K in a magnetic field of 20 mT, i.e. through the skyrmion phase down to 12 K, the lowest temperature measured in this orientation. The cooling process took about 8 minutes. The desired magnetic field was subsequently applied and the SANS diffraction patterns were recorded upon heating in small temperature steps. In this way metastable skyrmions were nucleated at temperatures below the skyrmion range [34-38]. This manifests in the coexistence of the helical and metastable skyrmion phase. At low magnetic fields a diffraction pattern with only a fourfold symmetry is observed. This is an indication for the helical phase which is twinned with a propagation direction along the two in-plane cubic crystal axis. Above 10 mT an additional structure with a sixfold symmetry develops in the diffraction pattern (see Fig. 3.i). This indicates the coexistence of the helical and the metastable skyrmion phases, which exists over a broad range in the H-T phase diagram from the LT-SkL to the HT-SkL phase. This coexistence is absent for the ZFC data (see Fig. 3.c for ZFC in comparison to Fig. 3.i for FC at 20 mT). It is interesting to note that, in case of this coexistence the absolute |q|-value for the helical propagation vector and the skyrmion lattice are perfectly identical. The 360° circular integration of the neutron scattering intensity I(q) round the centre of the neutron beam for the individual diffraction pattern is also shown in Fig. 3. The obtained peaks were analysed by fitting Gaussian lineshapes to the experimental data. The obtained result for the helical-SkL coexistence as shown in Fig. 3.i does not indicate any peak splitting. Even the linewidth remains identical when compared to the linewidth obtained for the pure helical diffraction pattern as obtained after ZFC at the same temperature and magnetic field (see Fig. 3.c). The absence of a peak splitting or broadening in the circularly integrated data is also an evidence that both, the propagation vector of the helical phase and the lattice parameters of the SkL phases (HT and LT) are perfectly isotropic along the two in-plane ⟨1 0 0⟩ cubic directions. The same holds for the unoriented sample (see Supplemental Fig. S2) and the ⟨1 1 0⟩ oriented sample (See Supplemental Fig. S4), where the two in-plane directions are along ⟨1 0 0⟩ and ⟨1 1 0⟩. This is in contrast to our observation on thinned COSO single crystals where the propagation vector of the helical phase along the ⟨1 0 0⟩ direction had a different value and magnetic field dependence as compared to the ⟨1 1 0⟩ direction [16].

Finally, the sample was oriented with the ⟨1 1 0⟩ cubic crystal axis parallel to the magnetic field and neutron beam. The data shown in Figs. 4.a) and b) were taken after varied field cooling (VFC). In this case the desired magnetic field was adjusted in the paramagnetic phase at 80 K to the field in which the sample will be measured and the sample was then cooled down to the lowest possible temperature. The diffraction patterns were recorded upon heating in small temperature steps and selected diffraction maps are shown in the Supplementary Information (Fig. S4). Using this procedure, the HT-SkL phase is passed when cooling below 30 mT, leading to the nucleation of metastable skyrmions. Strong Bragg reflections of the helical phase are observed below 20 mT. Coexisting helical and skyrmion features are only visible between 20 mT and 30 mT. Again, their absolute |q|-values are identical and the Bragg peaks appear perfectly isotropic in the $q_x$-$q_y$ plane of the detector. In contrast to the data taken on the unoriented sample, weak Bragg reflections of the helical phase are visible up to higher magnetic fields, i.e. beyond 37 mT at 14 K and the helical and HT-SkL phase are not separated by the conical phase. Furthermore, there is no evidence for a LT-SkL phase as observed for the ⟨1 0 0⟩ crystal orientation. Concerning the |q|-value the main features seen for the unoriented single crystal are reproduced, i.e. |q| possesses a local maximum right in the centre of the HT-SkL phase and the maximum |q|-value of the helical propagation vector is reached at zero magnetic field between about 35 K to 55 K. In contrast to the unoriented crystal at ZFC where the conical phase separates the helical and the HT-SkL phase, the |q|-value can be determined throughout the coexistence range of the helical and metastable SkL phases.

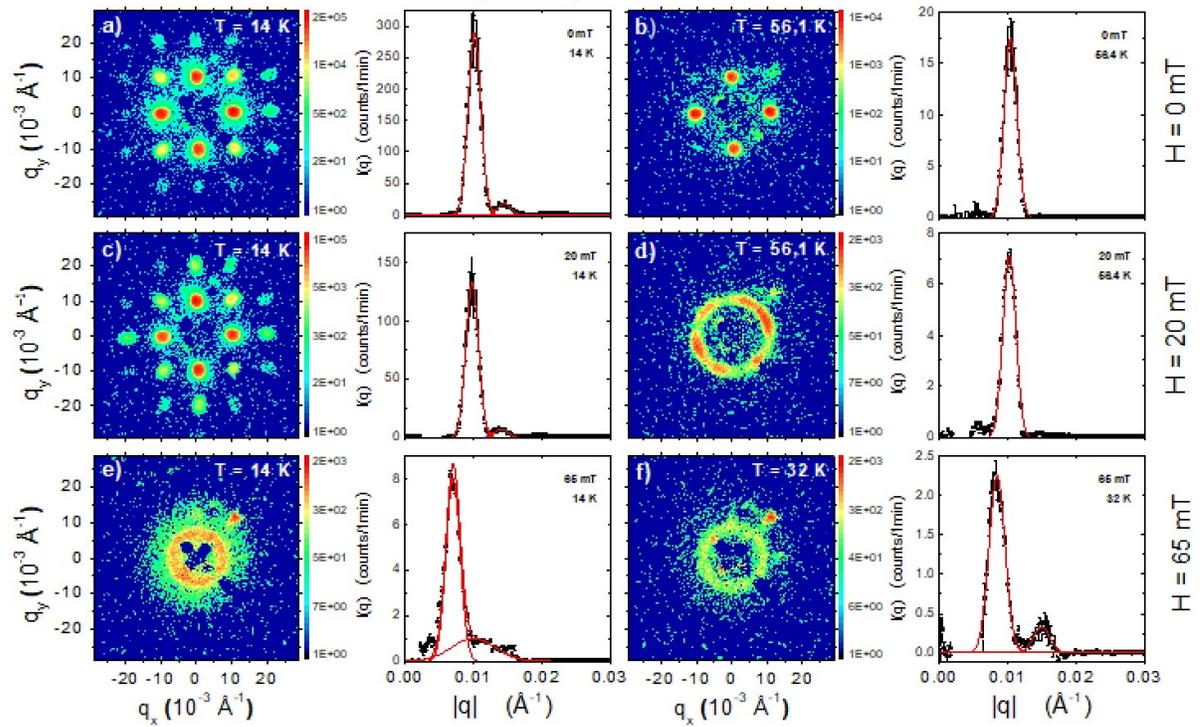
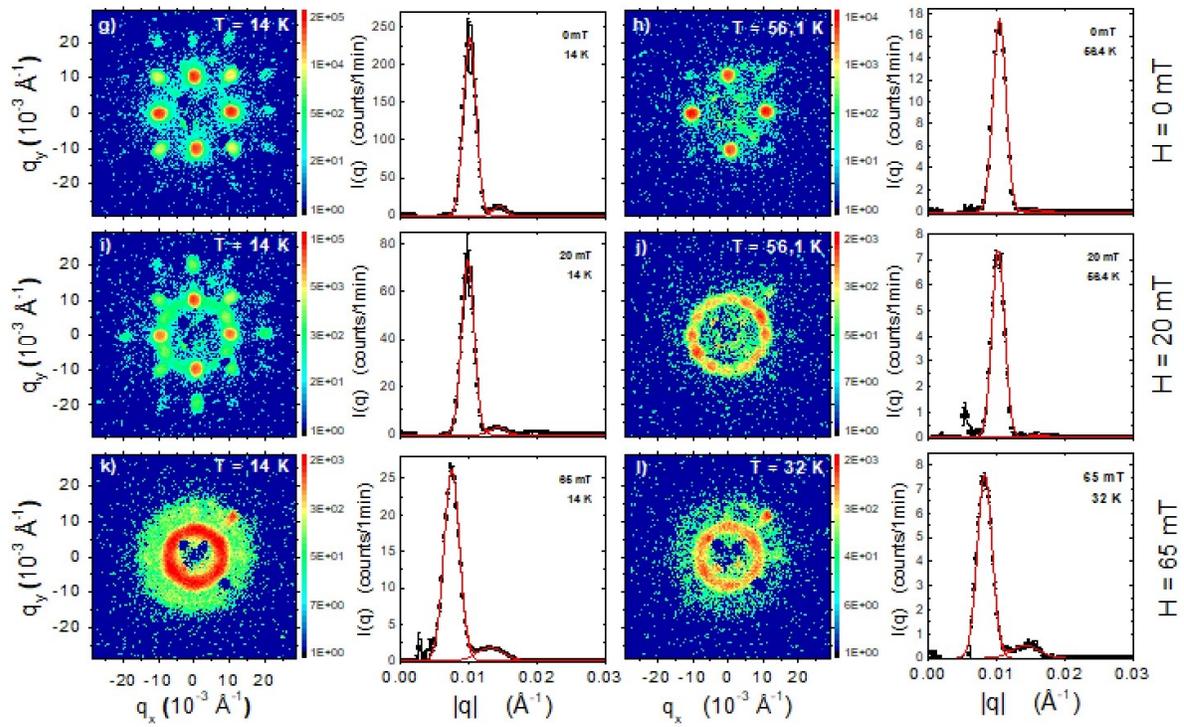

*Fig. 3: SANS diffraction patterns of the ⟨1 0 0⟩ oriented sample taken at different temperatures and magnetic fields. The upper part a) - f) displays the data taken after zero field cooling while the lower panel g) – l) shows the data taken after cooling from the paramagnetic phase at a magnetic field of 20 mT, i.e. by cooling through the skyrmion phase. In this case a coexistence of the helical phase and a metastable skyrmion phase is present and the HT- and LT-skyrmion phases exhibit higher Bragg peak intensities (see lower panels j) to l)). Next to the diffraction maps the results of the circular integration around the centre of the neutron beam are shown. The red lines are the Gaussian lineshapes which were fitted to the experimental data. The error bars correspond to 1 s. d. of the intensity I(q).*

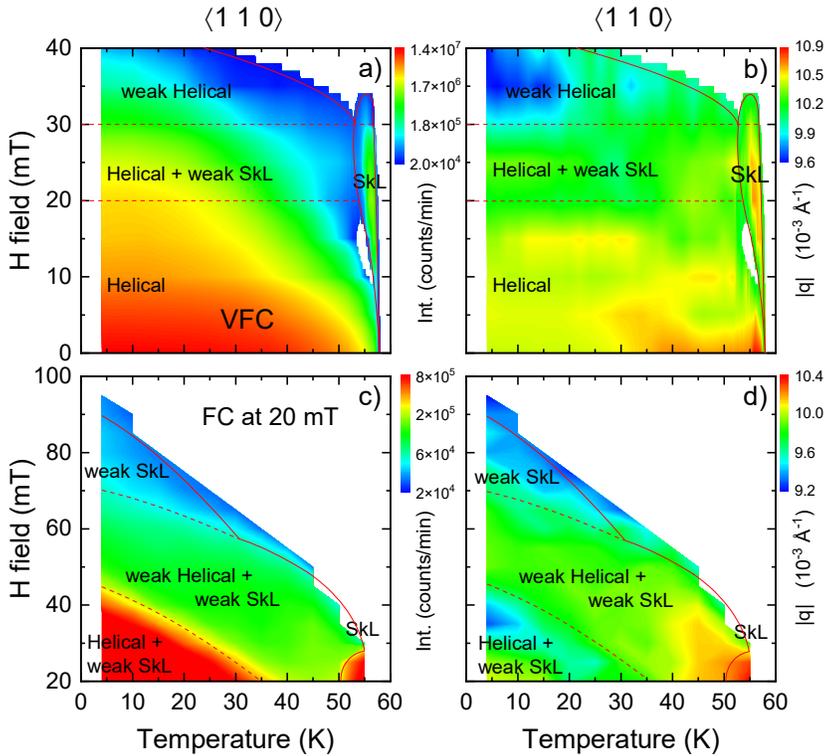

*Fig. 4: H-T phase diagram of the sample with the ⟨1 1 0⟩ cubic crystal axis parallel to the magnetic field and neutron beam. The upper panels a) and b) show the data taken after fixed field cooling, where the magnetic field was fixed to a certain value before cooling from 80 K. For the lower panels c) and d) the data were taken after field cooling at 20 mT, i.e. through the skyrmion phase. The contour maps of the logarithmic intensities are shown in a) and c) while the |q|-values are shown in b) and d). The red solid lines serve as guides to the eye.*

The HT-SkL has a smaller stability range as compared to the unoriented crystal. The transition to the paramagnetic phase is at 57.5 ± 0.15 K at 25 mT as compared to 58.2 ± 0.15 K for the unoriented single crystal and 57.0 ± 0.15 K for the ⟨1 0 0⟩ crystal orientation measured for both cooling methods, ZFC and FC at 20 mT (see Supplementary Fig. S3). This is right in between the values of the unoriented and the ⟨1 0 0⟩ oriented crystal and serves as a further indication that the stability range of the HT-SkL is controlled by the magnetic anisotropy term, which is strongest for the ⟨1 0 0⟩ crystal orientation. The lower boundary of the HT-SkL is 54.6 ± 0.15 K at 25 mT, which is comparable to the value of 54.2 ± 0.15 K obtained for the unoriented sample. It must be noted that the data of the ⟨1 1 0⟩ oriented crystal were taken after VFC, i.e. metastable skyrmions were nucleated.

Figure 4 c) and d) shows the intensities and corresponding |q|-values for the crystal oriented along the ⟨1 1 0⟩ cubic crystal axis after FC at 20 mT. After cooling through the HT-SkL phase to the final temperature the data were recorded upon increasing the magnetic field from 20 mT in 5 mT steps. At 20 mT the coexistence of strong Bragg peaks of the helical and weak peaks of the HT-SKL phase is observed at temperatures up to 55 K and the HT-SkL phase exists between 55 K and 57.5 K. The diffraction patterns are shown in the Supplementary Information (Fig. S4). A coexistence of the helical and metastable skyrmion phase is present between the HT-SkL phase and up to 70 mT at 4 K. This is indicated in both the intensity (Fig. 4.c) and the |q|-value (Fig. 4.d); however, at 4 K with magnetic fields above 70 mT, diffraction features of the skyrmion phase are still present up to about 95 mT. In this region the |q|-value reaches a minimum, i.e. the skyrmion distances reach their largest value as in the case for the LT-SkL phase observed for the ⟨1 1 0⟩ crystal orientation. Nonetheless, due to the six-fold Bragg peak pattern and not a ring-like diffraction structure, we would not attribute the SkL phase in this range to the LT-SkL phase but rather to metastable skyrmions which were nucleated when cooling through the HT-SkL phase.

Figure 5 presents the integrated intensities, absolute |q|-values and λ, which is either the length of the spin helical or the skyrmion distance in the hexagonal skyrmion lattice. The HT-SkL is shaded as red areas. In case of field cooling (either VFC or FC at 20 mT) metastable skyrmions are nucleated which coexist with the helical spin structure. This is indicated by the yellow areas. The blue areas show the LT-SkL. The intensity of the LT-SkL Bragg peaks has a flat plateau at low temperature while the HT-SKL has a sharp maximum at about 57 K before melting upon heating. This suggests that the LT-SkL phase does not strongly depend on the

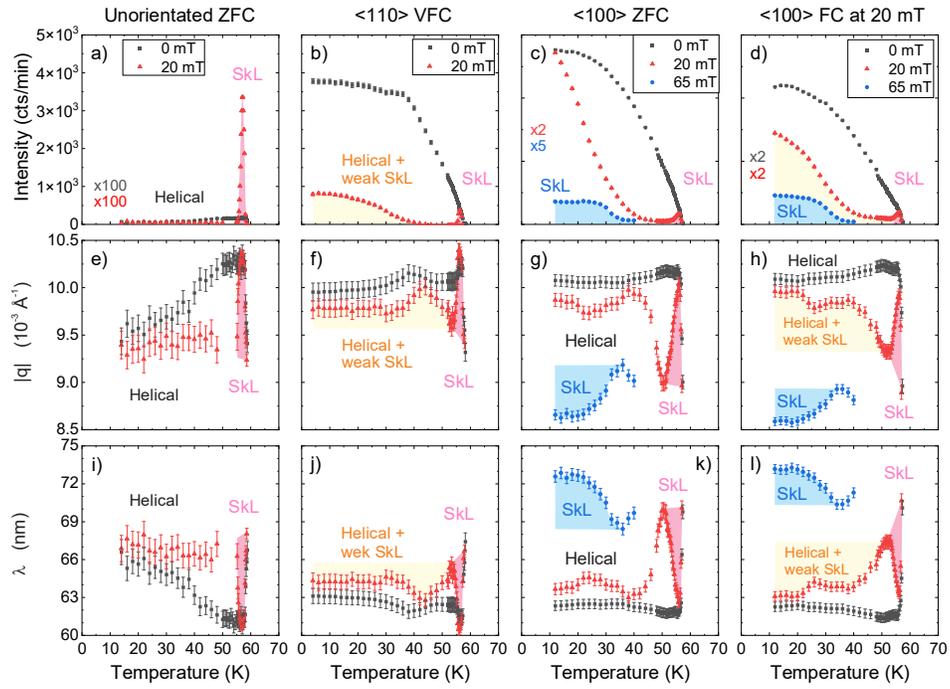

*Fig. 5: Temperature dependence of the integrated intensities, the absolute |q|-values and λ, i.e. the length of the spin helix or the skyrmion distances, respectively, at selected magnetic fields, i.e. 0 mT, 20 mT and 65 mT. Shown are the data taken after ZFC of the unoriented single crystal, VFC of the ⟨1 1 0⟩ oriented crystal, and after ZFC and FC at 20 mT of the ⟨1 0 0⟩ oriented single crystal. The red areas represent the HT-SkL phase and the blue areas the LT-SkL phase. The yellow areas indicate the range where the helical and metastable skyrmion phases coexist. The error bars correspond to 1 s. d. of the fits of Gaussian lineshapes to the experimental data.*

sample temperature. The skyrmion distance reaches a minimum in the centre of the HT-SkL, i.e. they drop from about 67.0 nm at the boundaries to about 60.5 nm for both the unoriented and the ⟨1 1 0⟩ oriented crystal, and from about 70.0 nm at the boundaries to 63.0 nm for the ⟨1 0 0⟩ oriented crystal for both cooling conditions ZFC and FC at 20 mT. The LT-SkL phase is only established for the ⟨1 0 0⟩ oriented crystal. In contrast to the HT-SkL, the skyrmion distances reach a flat maximum of about 73.0 nm at 12 K and decrease with increasing temperature to about 70.5 nm at 40 K. The inverse behaviour of the skyrmion distances for the HT-SkL and LT-SkL indicates that both skyrmion phases are not connected to each other and originate from the contributions of different terms in the spin Hamiltonian. It is also important to note that the unoriented and the ⟨1 1 0⟩ oriented crystal possess almost the same skyrmion distances at the boundaries and the centre of the skyrmion phase, but the skyrmion distances increases for the ⟨1 0 0⟩ oriented single crystal. On the other hand, the different cooling methods of ZFC and FC at 20 mT have almost no effect on the skyrmion distance for the ⟨1 0 0⟩ oriented single crystal. An important observation is the fact that λ is continuous at the transition from the helical to the HT-skyrmion phase. This holds for both cases, after ZFC where only the helical phase is established and, after FC, when both the helical spin structure and metastable skyrmions coexist. This information will provide important input for the further understanding of the stabilization of the skyrmion phases; in particular, which terms in the spin Hamiltonian play a dominating role for the formation of the HT-SkL and LT-SkL phases.

**DISCUSSION**

Detailed investigations of the scaling behaviour of the length of the helical propagation vector and skyrmion distances have been performed in the multiferroic skyrmion material $Cu_2OSeO_3$. Therefore, the crystal was mounted in various orientations along the magnetic field and incident neutron beam, i.e. unoriented, oriented along the ⟨1 0 0⟩ and along the ⟨1 1 0⟩ crystal directions. Furthermore, various cooling protocols were examined, zero field cooling and field cooling (at 20 mT and at an initial fixed magnetic field upon cooling), which resulted in the nucleation of metastable skyrmions before starting with the measurements. SANS neutron diffraction patterns were recorded at various magnetic fields and temperatures in order to map out the Bragg peak intensities and peak positions of the various magnetic phases throughout the different H-T phase diagrams. The absolute |q|-value of the Bragg peak positions was determined by performing a circular integration of the Bragg peak intensities around the centre of the incident neutron beam and then fitting of Gaussian lineshapes to the experimental data. Using this method, a detailed picture of the stability range of the high-temperature and low-temperature skyrmion phases as well as the helical phase was obtained.

A systematic reduction of the phase transition temperature from the HT-SkL to the paramagnetic phase was observed upon crystal orientation from the unoriented to the ⟨1 1 0⟩ oriented and finally to the ⟨1 0 0⟩ oriented crystal. The full H-T phase diagram of the absolute |q|-value of the helical propagation vector and the skyrmion lattice provided valuable insight

into the mechanisms for the formation and stability of the skyrmion phases in COSO. For the ⟨1 0 0⟩ oriented crystal two skyrmion phases appear, a HT-SkL and LT-SkL phase. For the HT-SkL the skyrmion distances λ reach a minimum in the centre of the skyrmion range while for the LT-SkL the skyrmion distances possess a broad maximum at lowest temperatures. This indicates that both skyrmion phases have distinct behaviour. Upon field cooling through the HT-SkL phase metastable skyrmions were nucleated. This results in a coexistence of the helical phase and metastable skyrmions. It is important to note that both phases possess the identical |q|-value and therefore the same length scale throughout their coexistence in the entire phase diagram. Furthermore, λ is continuous at the transition from the helical to the HT-skyrmion phase irrespective as to whether metastable skyrmions were generated or not.

The obtained results provide important information for the theoretical understanding of the formation and stability of skyrmions. Furthermore, the data show how skyrmions, in particular their distances and therefore their density, can be manipulated by magnetic field and temperature. Knowledge about this tunability serves as an important input for future technological applications, for example in data storage devices in information technology, in spintronics or in skyrmion quantum computation.

**METHODS**

*Growth and characterization of COSO single crystals*

Single crystals of $Cu_2OSeO_3$ were grown by Chemical Vapor Transport (CVT) using a temperature gradient. A mixture of 0.511 g of CuO powder (Achtung Chempur 99.99%), 0.489 g of $SeO_2$ (Sigma Aldrich 99.99%) and 0.030 g of Ammonium Chloride $NH_4Cl$ (Analar BDH 99.5%) as transport agent was inserted into a quartz tube of approximately 10 cm of length evacuated and sealed with an $H_2/O_2$ flame. The mixture was then heated in a two-zone furnace with a source temperature of 883.15 K and a sink temperature of 813.15 K. The obtained single crystals were rinsed with ethanol and deionised water. The crystal chosen for this neutron scattering experiment has a size of about $3 \times 3 \times 3.2$ mm$^3$ and a picture of the crystal is shown in Supplementary Information (Fig. S1.a). To verify the quality of the single crystal and to align the crystal in different orientations, i.e. with the incident neutron beam along the ⟨1 0 0⟩ and ⟨1 1 0⟩ crystal axis, neutron Laue diffraction experiments were performed using the instrument JOEY at the OPAL research reactor of the Australian Nuclear Science Technology Organization (ANSTO), Australia. The diffraction patterns representing the different orientations are shown in the Supplementary Information (Figs. S1.b-d). An aperture of 4 mm diameter ensured that the crystal was fully illuminated and the absence of double reflections confirms that the obtained crystal was not twinned. In addition, magnetization measurements were performed using a Superconducting Quantum Interference Device (MPMS3, Quantum Design). The data was collected in vibrating-sample magnetometer mode as a function of the applied magnetic field at different temperatures. They serve as a further evidence for the excellent quality of the grown crystal.

*Small angle neutron scattering experiments*

Small angle neutron scattering (SANS) experiments were performed using the instrument QUOKKA at ANSTO [45-47]. A wavelength of $\lambda \sim 5 \pm 0.50$ Å was chosen using a neutron velocity selector. An area detector of 1 m x 1 m and a spatial resolution of 5 mm x 5 mm was placed a distance of 20 m behind the sample. In order to determine the full magnetic field – temperature phase diagram of the helical and skyrmion phases of COSO, the single crystal was placed inside a cryo-magnet to apply magnetic fields parallel to the incident neutron beam.

**ACKNOWLEDGEMENTS**

This work was supported through the Australian Research Council (ARC) through the funding of the Discovery Grant DP170100415 and the funding of the Linkage Infrastructure, Equipment and Facilities Grant LE180100109. O.A.T. acknowledges the support from the Australian Research Council (Grant No. DP200101027) and the NCMAS grant. Furthermore, this work was supported through the New Zealand's Marsden Fund 20-UOA-225. We thank the Australian Centre for Neutron Scattering at ANSTO for the allocation of neutron beam time on the instruments JOEY and QUOKKA.

**AUTHOR CONTRIBUTIONS**

J.S.F., R.R., J.O'B., S.Y., M.F.P., M.S., J.V., N.B., E.G. and C.U. participated in the neutron experiments. J.S.F and C.U. performed the data analysis. R.R., S.Y., M.S., J.V., and C.U. prepared and characterized the samples. O.T. perform the theoretical modelling. T.S. and C.U. conceived and supervised the project. J.S.F and C.U. wrote the manuscript. All authors contributed to the manuscript and interpretation of the data.

ADDITIONAL INFORMATION

Supplementary Information is attached below.

SUPPLEMENTARY INFORMATION

# Scaling behaviour of the helical and skyrmion phases of $Cu_2OSeO_3$ determined by single crystals small angle neutron scattering

J. Sauceda Flores,[1] R. Rov,[2] J. O'Brien,[1] S. Yick,[1,2,3,4] Md. F. Pervez,[1] M. Spasovski,[2] J. Vella,[2] N. Booth,[4] E. Gilbert,[4] Oleg A. Tretiakov,[1] T. Söhnel[2,3,†] and C. Ulrich[1,*]

[1] *School of Physics, The University of New South Wales, NSW 2052, Australia*
[2] *School of Chemical Sciences, University of Auckland, Auckland 1142, New Zealand*
[3] *MacDiarmid Inst. Adv. Mat. & Nanotechnology, Wellington 6140, New Zealand*
[4] *Australian Centre for Neutron Scattering, Australian Nuclear Science and Technology Organisation, Lucas Heights NSW 2234, Australia*

(Dated: 4. March 2023)

[†] Corresponding Author: t.soehnel@auckland.ac.nz
[*] Corresponding Author: c.ulrich@unsw.edu.au

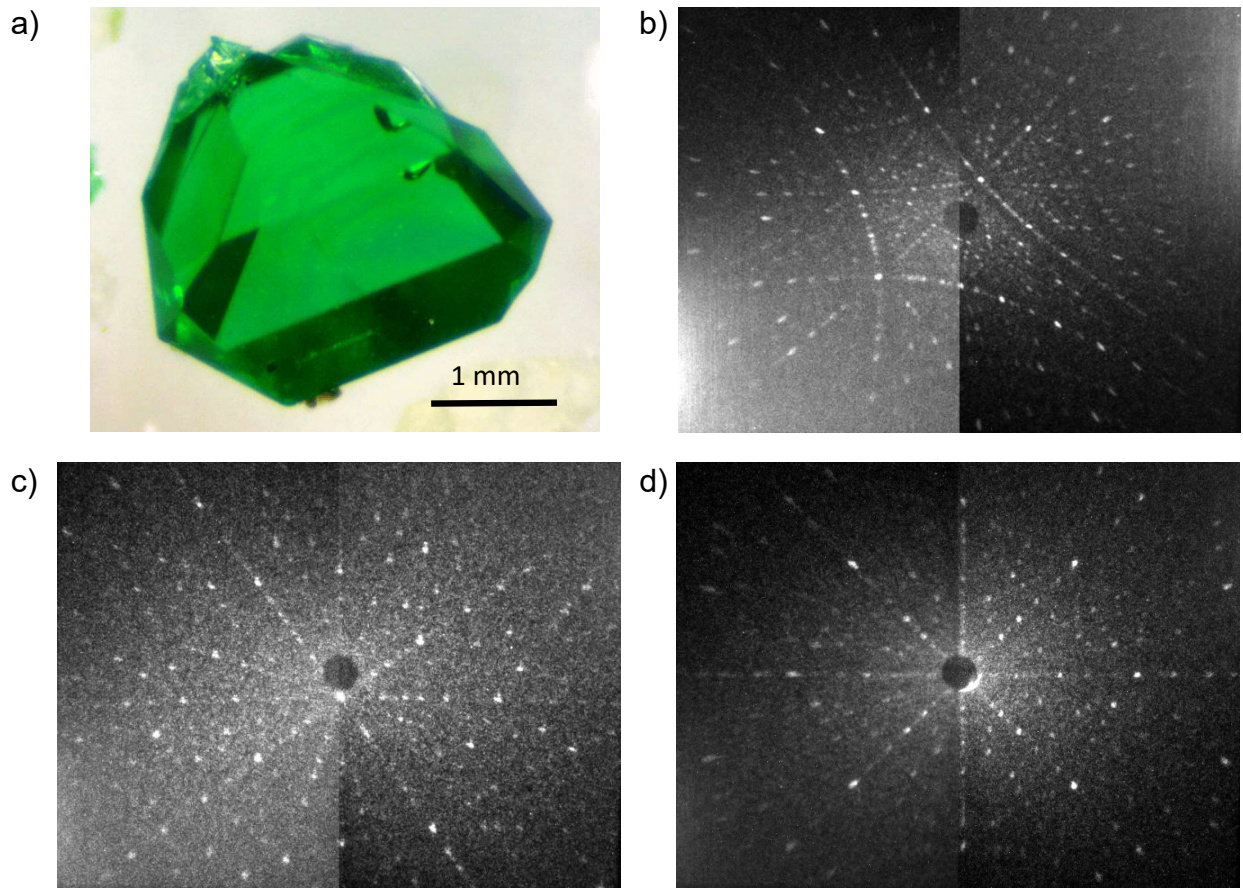

**Figure S1:** a) Picture of the $Cu_2OSeO_3$ single crystal used for the small angle neutron scattering experiments. b)-d) Neutron Laue diffraction of the $Cu_2OSeO_3$ single crystal measured on the instrument JOEY at the ANSTO. The patterns were recorded with an illumination time of 300 sec using an aperture of 4 mm to ensure a full illumination of the crystal. The data demonstrate the high quality of the single crystal and the absence of twining within the instrumental resolution. b) Neutron Laue diffraction image of the unoriented sample used for the SANS experiment, c) for the incident neutron beam oriented along the $\langle 1\ 1\ 0 \rangle$ crystal axis and d) for the $\langle 1\ 0\ 0 \rangle$ crystal axis along the incident neutron beam, respectively.

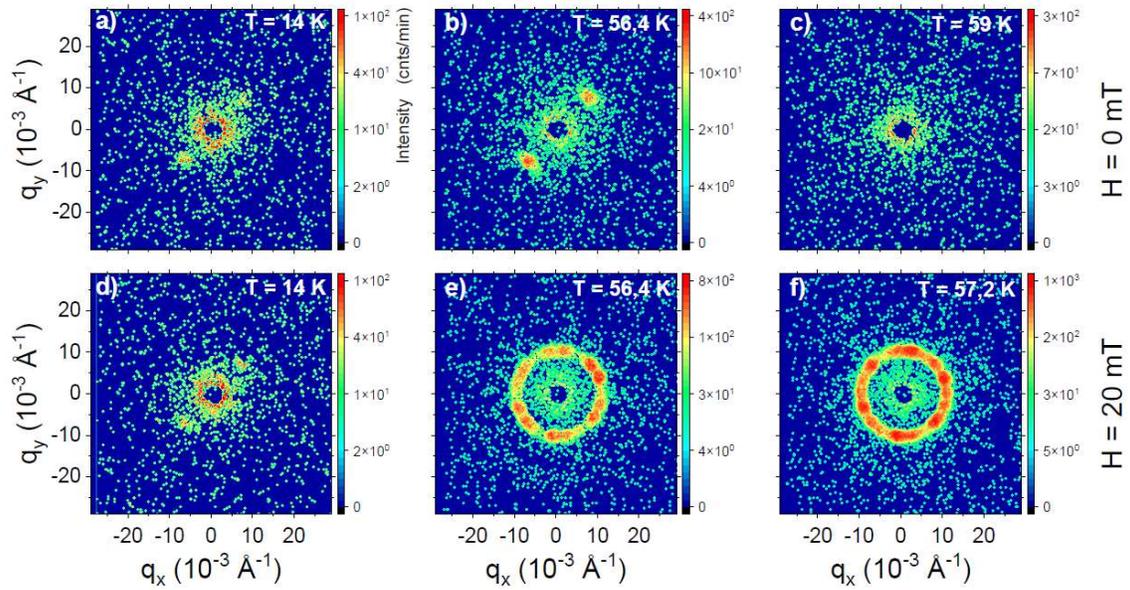

**Figure S2:** SANS diffraction patterns of the unoriented COSO single crystal taken at different temperatures at magnetic fields of 0 mT and 20 mT. a) and b) the data were taken at 0 mT and show the typical diffraction pattern of the helical phase with two Bragg reflections. c) data taken at 59 K, i.e. above the magnetic ordering temperature. The absence of any diffraction peaks confirms the paramagnetic phase. d) for 20 mT and T = 14 K the sample is well within the helical phase and the expected two Bragg reflections are observed. e) and f) correspond to the skyrmion phase. In total 12 Bragg reflections are observed, which indicates the existence of two skyrmion lattices, which are tilted by about 25° degree.

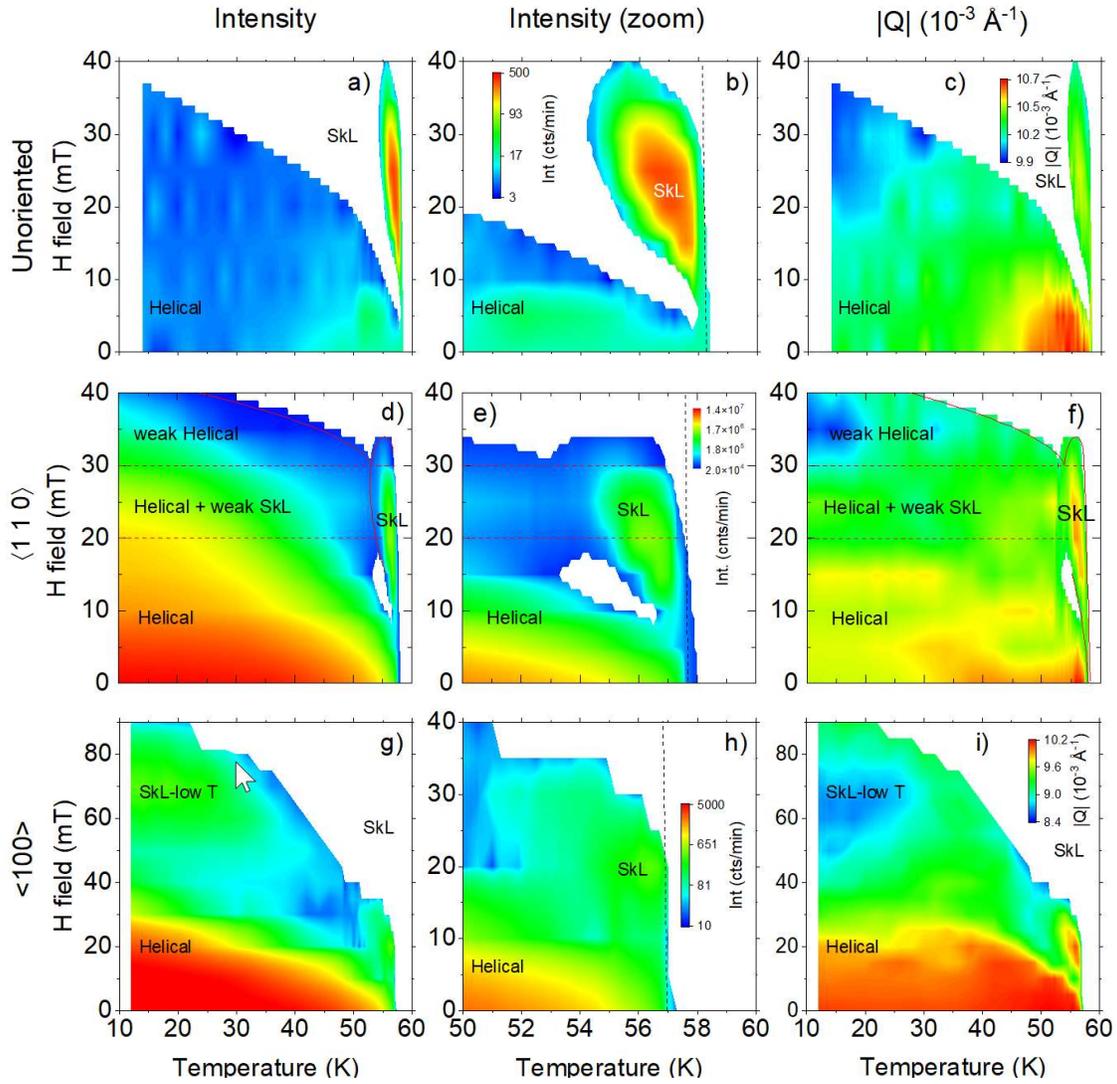

**Figure S3:** H-T phase diagram of the unoriented COSO single crystal (upper row), ⟨1 1 0⟩ oriented crystal (middle row) and ⟨1 0 0⟩ oriented crystal (lower row). The integrated intensities are shown in a), d) and g) and in a zoomed in range in b), e) and h) while the absolute |q|-values of the Bragg peaks are shown in c), f) and i). The data sets on the unoriented and the ⟨1 0 0⟩ oriented crystal were measured after ZFC while the H-T phase diagram of the ⟨1 1 0⟩ oriented crystal was taken after FC at 20 mT. The direct comparison of the high temperature skyrmion phase as shown in b), e) and h) indicates that boundary of the HT-SkL phase to the paramagnetic phase is reduced by 0.7 K for the ⟨1 1 0⟩ oriented crystal and lowered by 1.2 K for the ⟨1 1 0⟩ oriented crystal. The temperature of the lower boundary of the HT-SkL is comparable for the unoriented and ⟨1 1 0⟩ oriented crystal but shifted by 3.2 K to lower temperatures for the ⟨1 0 0⟩ oriented crystal.

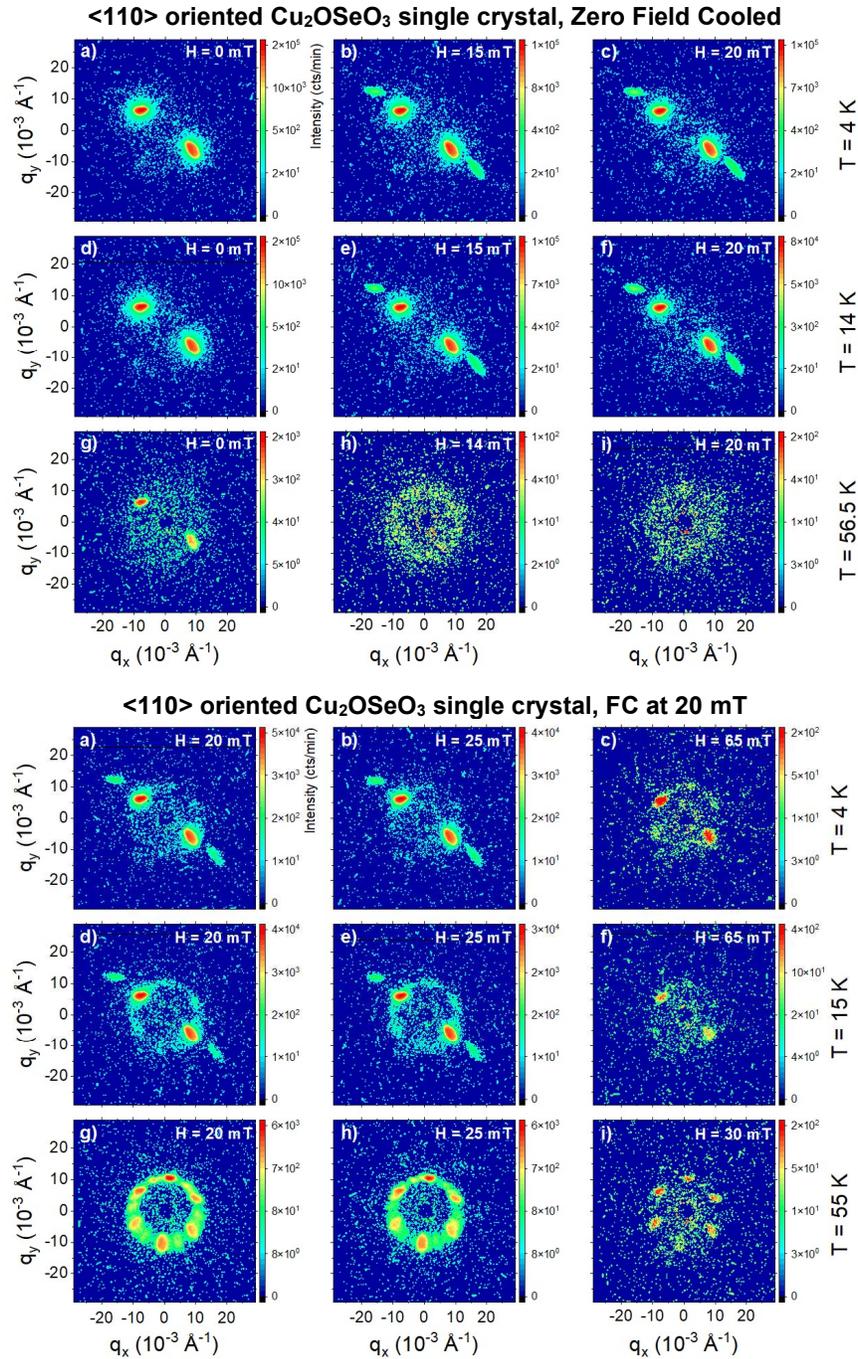

**Figure S4:** Comparison of the SANS diffraction patterns of the ⟨1 1 0⟩ oriented sample taken at different temperatures at magnetic fields. The upper panel displays the data taken after ZFC and the lower panel shows the data taken after FC at 20 mT, i.e. through the HT-SkL phase. In this case a coexistence of the helical phase and a metastable skyrmion phase is present (see lower panel a) to f)) and the skyrmion phase is better established (see lower panel g) to i)). Note, the small distortion of the Bragg peaks is a result of slight misalignment of the SANS instrument.